# Branch: An interactive, web-based tool for testing hypotheses and developing predictive models

Karthik Gangavarapu [1], Vyshakh Babji[1], Tobias Meißner[1,2], Andrew I. Su[1,*] and Benjamin M. Good[1,*]

[1]Department of Molecular and Experimental Medicine, The Scripps Research Institute, 10550 North Torrey Pines Road, La Jolla, CA, 92037 ., [2]Avera Health

*To whom correspondence should be addressed.

## Abstract
**Summary:** Branch is a web application that provides users with no programming with the ability to interact directly with large biomedical datasets. The interaction is mediated through a collaborative graphical user interface for building and evaluating decision trees. These trees can be used to compose and test sophisticated hypotheses and to develop predictive models. Decision trees are evaluated based on a library of imported datasets and can be stored in a collective area for sharing and reuse.

**Availability and Implementation:** Branch is hosted at http://biobranch.org/ and the open source code is available at http://bitbucket.org/sulab/biobranch/.
**Contact:** {gkarthik, asu, bgood}@scripps.edu

## 1  Introduction

One central goal of large-scale molecular profiling is to identify consistent patterns that can be used to advance understanding of and to make predictions about a particular condition. Given the large numbers of features typically measured, it is assumed that complex patterns involving interactions between multiple variables exist and can be detected by modern statistical and machine learning approaches. Yet, despite the troves of data collected, the predictive and explanatory power of patterns extracted automatically from data remain weak for most conditions of interest(Weigelt *et al.*, 2012). Further, it ranges from challenging to impossible for biomedical domain experts to directly engage with the data. Apart from researchers with advanced programming and statistical skills at their disposal, it is nearly impossible to answer even simple questions from a breast cancer data set like:

1. "do the expression levels of VEGFA correlate with survival?"
2. "if AURKA expression is high and TOP2A expression is low, is the risk of recurrence lower or higher?"
3. "are the genes in the apoptosis pathway more or less predictive of survival than those involved in Mitochondrial complex I activity?"

Here we introduce an interactive Web application, called Branch, that makes it easy for anyone to answer questions like these. Branch can be used to test hypotheses and to construct and evaluate predictive models. It also provides a collaborative graphical interface for manual creation of decision trees that encapsulate the structure of sophisticated hypotheses. This tool not only allows users to answer questions of particular datasets, it can also be used to interactively construct and test a complex predictive model. Such models may, because of the incorporation of the user's expertise, outperform models inferred from the data by strictly automated methods (Stumpf *et al.*, 2009).

## 2  Using Branch

Users begin by selecting a dataset of interest from the Branch dataset library. Each dataset corresponds to a single table in which each row contains the values for a set of features, e.g. gene expression levels or clinical variables, and a binary class label for each sample, e.g. cancer/normal. The library currently contains several datasets selected to demonstrate the features of the application. Additional datasets may be loaded upon request. Branch may also be installed locally from its open source code.

Once a dataset is selected, the user chooses an evaluation option to be used to measure the quality of the decision tree they will build in the subsequent step. They may select from three options. First, selecting 'training set' allows users to see exactly how a tree fits the particular dataset under study. This may lead to false conclusions about the generality of the constructed tree ('overfitting'). Second, users can select 'test set' for cases where explicit training and test sets are provided. In this case, the user builds the tree with information from the training set but sees evaluation results on the test set. Third, if no compatible datasets are available as independent test sets, the 'percentage split' option simulates the process by dividing the data randomly into a training and testing set – again giving more realistic assessments about the generalizability of decision tree structures developed using the tool.

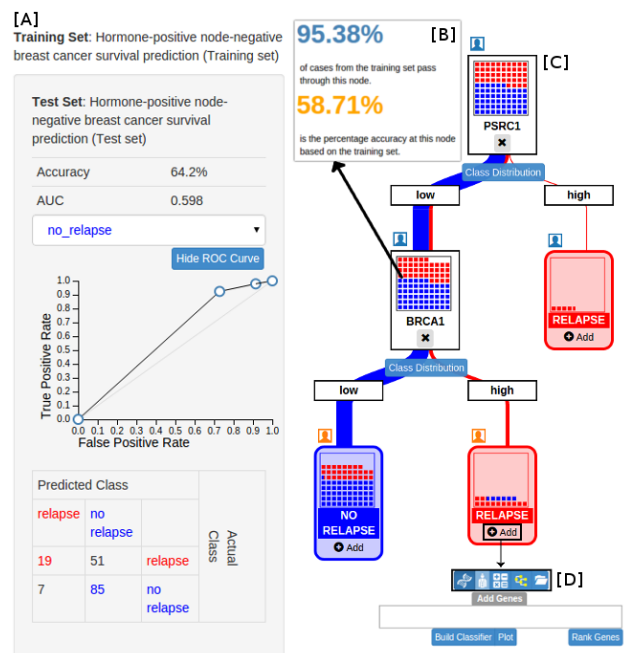

**Figure 1.** A decision tree built using a node-negative, ER-positive, HER2-negative breast cancer dataset(Griffith *et al.*, 2013). (A) Evaluation of the tree on the testing set. The evaluation sidebar shows the accuracy, area under the curve and the confusion matrix. (B) The percentage of samples with "low" expression of PSRC1 and the accuracy of the prediction at the concerned leaf node are shown in the dialogue box. (C) The decision tree as visualized in Branch. (D) The search bar used to add split nodes to the decision tree.

Given a dataset and an evaluation method, the user can begin constructing decision trees and measuring their quality (Figure 1). Building a tree corresponds to the process of iteratively adding split nodes. Branch supports five different split node types. Most simply, single feature splits may be created from individual features in the dataset such as the expression values for a particular gene or the age of a patient. Custom features may be created as linear combinations of other features. For example, the OncotypeDx (Paik *et al.*, 2004) breast cancer recurrence score can be recreated and applied as a feature for use in single Branch split nodes (See supplementary data). The user may also choose to use built-in machine learning algorithms to infer a predictive model from a feature subset and use the model for a decision node. Likewise, previously created trees can be used as individual decision nodes. Finally, the system provides a visual split creator that lets the user define decision boundaries graphically (Ware *et al.*, 2001). A tree may incorporate mixtures of these different node types.

Users can begin building a tree from scratch or can select an existing tree that corresponds to their dataset from the community library or their personal collection. Once created, the user may save their tree to the public collection or keep it private.

## 3 Results

Branch provides a new mechanism to connect a large pool of biologically savvy (but perhaps not computationally savvy) researchers with large, high-dimensional datasets. In doing so, it should help them develop and refine better, more informed hypotheses. In addition, the application can be used to generate a crowdsourced collection of predictive models that captures the collective knowledge of the user community. This library of predictive models should prove useful to individual users and might also be applied computationally, for example, to generate an ensemble predictor.

Branch is available online at http://biobranch.org with open source code available at http://bitbucket.org/sulab/biobranch. It consists of a Java Spring server application that takes advantage of the Weka (Frank *et al.*, 2004) machine learning library and a Web client application based on Backbone.js and d3.js.

## Acknowledgements
We would like to thank Obi Griffith and Ryan Morin for discussions leading to the conceptualization of Branch and for testing early prototypes.

## Funding
This work was supported by NIH grants GM114833 and TR001114.

*Conflict of Interest:* none declared.

## References

Frank,E. *et al.* (2004) Data mining in bioinformatics using Weka. *Bioinformatics*, **20**, 2479–81.

Griffith,O.L. *et al.* (2013) A robust prognostic signature for hormone-positive node-negative breast cancer. *Genome Med.*, **5**, 92.

Paik,S. *et al.* (2004) A multigene assay to predict recurrence of tamoxifen-treated, node-negative breast cancer. *N. Engl. J. Med.*, **351**, 2817–26.

Stumpf,S. *et al.* (2009) Interacting meaningfully with machine learning systems: Three experiments. *Int. J. Hum. Comput. Stud.*, **67**, 639–662.

WARE,M. *et al.* (2001) Interactive machine learning: letting users build classifiers. *Int. J. Hum. Comput. Stud.*, **55**, 281–292.

Weigelt,B. *et al.* (2012) Challenges translating breast cancer gene signatures into the clinic. *Nat. Rev. Clin. Oncol.*, **9**, 58–64.